\documentclass{article}

\usepackage{microtype}
\usepackage{graphicx}
\usepackage{subfigure}
\usepackage{booktabs} 
\usepackage{bbding} 
\usepackage{multirow}
\usepackage{array} 
\usepackage{tabularx}
\usepackage{stfloats}

\usepackage{hyperref}

\newcommand\norm[1]{\lVert#1\rVert}

\usepackage[accepted]{icml2024}

\usepackage{amsmath}
\usepackage{amssymb}
\usepackage{mathtools}
\usepackage{amsthm}

\usepackage[capitalize,noabbrev]{cleveref}
\usepackage{comment}
\theoremstyle{plain}

\theoremstyle{definition}

\theoremstyle{remark}

\usepackage[textsize=tiny]{todonotes}

\icmltitlerunning{CLIPZyme: Reaction-Conditioned Virtual Screening of Enzymes}

\begin{document}

\twocolumn[
\icmltitle{CLIPZyme: Reaction-Conditioned Virtual Screening of Enzymes}



\icmlsetsymbol{equal}{*}

\begin{icmlauthorlist}
\icmlauthor{Peter G. Mikhael}{equal,mit}
\icmlauthor{Itamar Chinn}{equal,mit}
\icmlauthor{Regina Barzilay}{mit}
\end{icmlauthorlist}

\icmlaffiliation{mit}{Department of Electrical Engineering and Computer Science, Massachusetts Institute of Technology, Cambridge, MA, U.S.A}

\icmlcorrespondingauthor{Peter G. Mikhael}{pgmikhael@csail.mit.edu}
\icmlcorrespondingauthor{Itamar Chinn}{itamarc@csail.mit.edu}

\icmlkeywords{Machine Learning, ICML, enzyme screening, contrastive learning, biocatalysis representation learning}

\vskip 0.3in
]



\printAffiliationsAndNotice{\icmlEqualContribution} 

\begin{abstract}

Computational screening of naturally occurring proteins has the potential to identify efficient catalysts among the hundreds of millions of sequences that remain uncharacterized. Current experimental methods remain time, cost and labor intensive, limiting the number of enzymes they can reasonably screen. In this work, we propose a computational framework for \textit{in-silico} enzyme screening. Through a contrastive objective, we train CLIPZyme to encode and align representations of enzyme structures and reaction pairs. With no standard computational baseline, we compare CLIPZyme to existing EC (enzyme commission) predictors applied to virtual enzyme screening and show improved performance in scenarios where limited information on the reaction is available (BEDROC$_{85}$ of 44.69\%). Additionally, we evaluate combining EC predictors with CLIPZyme and show its generalization capacity on both unseen reactions and protein clusters.   

\end{abstract}

\section{Introduction}
\label{introduction}
Biosynthesis is the method of choice for the production of small molecules due to the cost effectiveness, scalability and sustainability of enzymes \citep{bornscheuer2012engineering, hossack2023building}. To find enzymes that can catalyze reactions of interest, practitioners often begin by identifying naturally occurring enzymes to repurpose or optimize. Only 0.23\% of UniProt is well studied and efficient enzymes likely lie among the hundreds of millions of sequences that are yet to be explored \citep{ribeiro2023enzyme}. As a result, the ability to computationally identify naturally occurring enzymes for any reaction can provide high quality starting points for enzyme optimization and has the potential to unlock a tremendous number of biosynthesis applications that may otherwise be inaccessible. 

In this work, we propose CLIPZyme, a novel method to address the task of virtual enzyme screening by framing it as a retrieval task. Specifically, given a chemical reaction of interest, the aim is to obtain a list of enzyme sequences ranked according to their predicted catalytic activity. In order to identify reaction-enzyme pairs, methods must contend with several unique challenges. First, in some cases, small changes in enzyme structures can lead to a large impact on its activity. Yet in other cases, multiple enzymes with completely different structural domains catalyze the same exact reaction \citep{ribeiro2023enzyme}. Similar principles hold for changes to the molecular structures of the reactants (substrates). This makes the task particularly challenging as methods must capture both extremes. Second, the efficacy of an enzyme is intricately linked to its interaction with the reaction's transition states \citep{marti2004theoretical, liu2021large}, which are difficult to model. Finally, in addressing the challenge of screening extensive datasets of uncharacterized enzymes, the scalability of computational methods becomes a critical factor.

CLIPZyme is a contrastive learning method for virtual enzyme screening. Originally developed to align between image-caption pairs, CLIP-style training has been successfully extended to model the binding of drugs and peptides to their target protein \citep{singh2023contrastive, palepudesign}. Unlike binding, however, the need to achieve transition state stabilization makes enzymatic catalysis a more nuanced process (in fact, very strong binding may inhibit an enzyme). Therefore, in order to represent the transition state, we develop a novel encoding scheme that first models the molecular structures of both substrates and products then simulates a pseudo-transition state using the bond changes of the reaction. To leverage the 3D organization of evolutionarily conserved enzyme domains, we encode AlphaFold-predicted structures \citep{jumper2021highly, varadi2022alphafold}. Since enzyme embeddings can be precomputed efficiently, screening large sets of proteins sequences for a new query reaction is computationally feasible. 

Since no standard method currently exists for virtual enzyme screening, we utilize enzyme commission (EC) number prediction as a baseline. Specifically, the EC number is an expert-defined classification system that categorizes enzymes according to the reactions they catalyze. Each EC number is a four-level code where each level provides progressively finer detail on the catalyzed reaction. For this reason, if a novel reaction is associated with an EC class, EC predictors can be used to identify candidate enzymes matching that EC class. 

We establish a screening set of 260,197 enzymes curated from BRENDA, EnzymeMap and CLEAN \citep{chang2021brenda, heid2023enzymemap, yu2023enzyme}. In our evaluation, we adopt the BEDROC metric, as is standard for virtual screening, and set its parameter $\alpha=85$. This places the most importance on the first $\sim$10,000 ranked enzymes, which constitutes a reasonable experimental screening capacity. We compare CLIPZyme to CLEAN, a state-of-the-art EC prediction model, on the virtual screening task and showcase its superior performance. While CLIPZyme can perform virtual screening without any expert annotations of reactions, methods like CLEAN cannot. We show that even when given some knowledge of a novel reaction's EC class, CLIPZyme is still superior to EC prediction for virtual screening (BEDROC$_{85}$ of 44.69\% compared to 25.86\%). Additionally, we show that combining CLIPZyme with EC prediction consistently achieves improved results. We also demonstrate that our reaction encoding outperforms alternative encoding schemes. Finally, we test our method on both unannotated reactions in EnzymeMap and a dataset of more challenging reactions involving terpene synthases \citep{samusevich2023terpene}.

\section{Related Work}

\paragraph{Reaction representation learning}
Methods to encode chemical reactions have been developed for a range of different computational tasks. This includes language models operating on reaction SMILES strings \citep{weininger1988smiles, schwaller2021extraction} and graph-based methods operating on the individual molecular structures of a reaction or on the condensed graph representations \citep{jin2017predicting, fujita1986description, hoonakker2011condensed}. These have shown strong performance on tasks like reaction rate prediction and forward synthesis \citep{madzhidov2014structure, heid2021machine}, but fail to take advantage of the data to effectively learn transition state representations. Models developed explicitly for transition state prediction are trained on simulations of very small molecules and are not scalable to enzymatic reactions \citep{duan2023accurate, van2023equireact}. In contrast to existing approaches that deterministically featurize bond changes, our method learns the features of these transition states directly from the data.


\paragraph{Catalysis of novel reactions} 
Successful design of enzymes most often begins with finding natural proteins that can subsequently be repurposed or optimized \citep{seelig2007selection, siloxanes2024sarai}. 
One option is to use EC prediction to filter enzyme screening sets. However, EC numbers are predefined by experts and provide a relatively coarse characterization of enzymes. As a result, one EC can capture many different reactions, while none may be able to capture a completely novel reaction. Therefore, filtering large libraries of enzymes by EC may yield impractically large sets of enzymes or none at all. Lastly, state-of-the-art EC predictors still show limited success (top F1 scores of 0.5-0.6) \citep{ayreshifi, yu2023enzyme, ryu2019deep, sanderson2023proteinfer}. In this work, we move away from human-crafted enzyme classes and instead operate directly on molecular and protein structures. 

Alternatively, the rational design of a new enzyme or active site requires a thorough understanding of the underlying mechanism \citep{rothlisberger2008kemp, jiang2008novo, yeh2023novo, feehan2021machine, weitzner2019computational}. While methods for protein sequence and structure generation have shown promise in creating custom folds and strong binders \citep{watson2023novo, ingraham2019generative, dauparas2022robust}, unnatural enzymes still suffer from low activity relative to naturally occurring ones \citep{hossack2023building}. Instead, we focus on identifying natural protein leads that can be optimized further either computationally or experimentally  \citep{seelig2007selection, bornscheuer2012engineering, siloxanes2024sarai}.


\section{Method}

\begin{figure*}[ht]
\vskip 0.1in
\begin{center}
\centerline{\includegraphics[width=\textwidth]{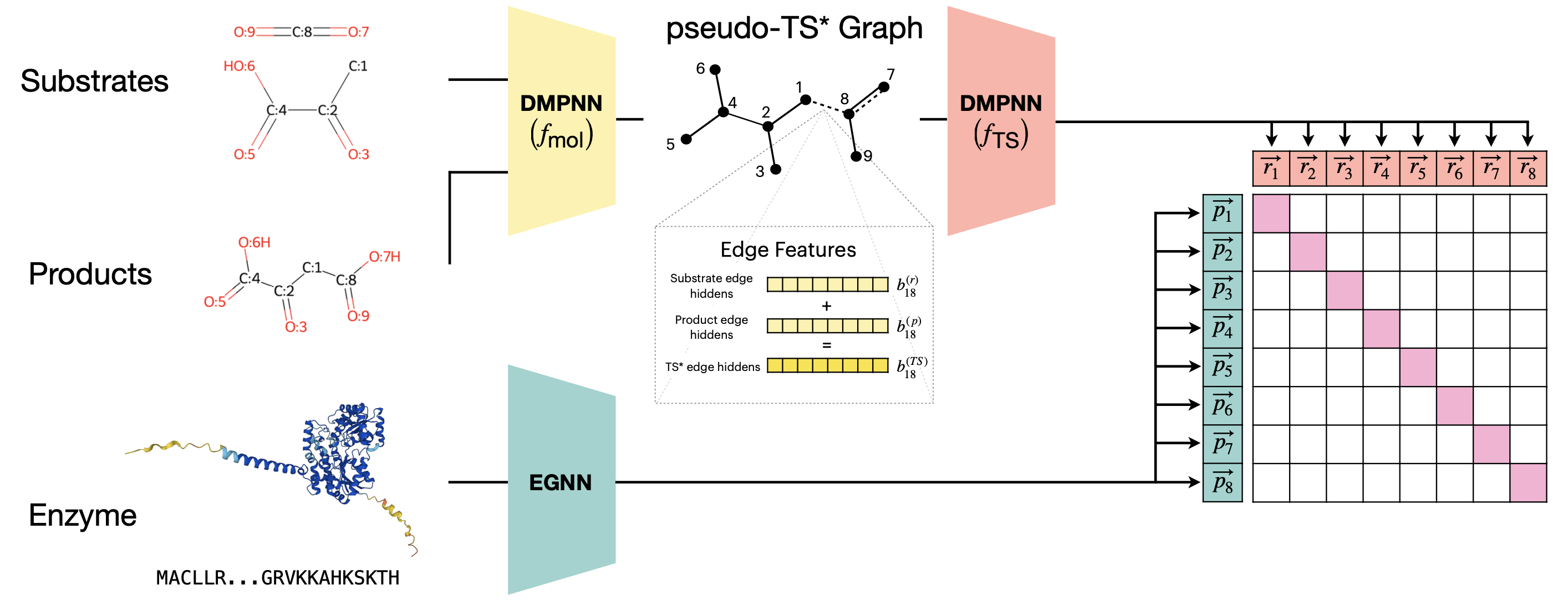}}
\caption{Overview of our approach. We encode atom-mapped chemical reactions using a DMPNN. We combine the substrate and product graphs by adding the hidden embeddings of their corresponding bonds to obtain an intermediate graph representing a pseudo transition state. A second DMPNN computes an embedding for the entire reaction. Enzymes are encoded with an EGNN using their predicted crystal structure and ESM-2 sequence embeddings. The reaction and enzyme representations are aligned with a CLIP objective.}
\label{fig:method-diagram}
\end{center}
\vskip -0.2in
\end{figure*}

We formulate enzyme screening as a retrieval task, where we have access to a predefined list of proteins and are asked to order them according to their ability to catalyze a specific chemical reaction. The representation of a protein $P$ is denoted by $p  \in \mathbb{R}^d$ and the query reaction $R$ by $r \in \mathbb{R}^d$. We aim to learn a scoring function $s(r, p)$ such that a higher score corresponds to a higher likelihood that $P$ catalyzes $R$.  We jointly learn a reaction encoder, $f_{rxn}$, and a protein encoder, $f_p$, to compute $r$ and $p$ (\cref{fig:method-diagram}). We adopt a contrastive learning objective \citep{sohn2016improved, radford2021learning} to maximize the cosine similarity between the embeddings of biochemical reactions and their associated enzymes (\cref{eq1,eq2}). We treat all enzymes in a training batch that are not annotated to catalyze a reaction as negative samples. Implementations details are provided in \cref{app:implementation,app:training}.
\begin{align}
    \label{eq1} s_{ij} &= s(r_i, p_j) = \frac{r_i}{\norm{r_i} }\cdot  \frac{p_j}{\norm{p_j} } \\
    \label{eq2} \mathcal{L}_{ij} &= -\frac{1}{2N} \left( \log \frac{ e^{s_{ij} / \tau} }{ \sum_i e^{s_{ij} / \tau} } + \log \frac{ e^{s_{ij} / \tau} }{ \sum_j e^{s_{ij} / \tau} } \right )
\end{align}
\subsection{Chemical Reaction Representation}
\label{sec:rxn-enc}
To obtain a functionally meaningful representation of the reaction, we leverage the key insight that the active sites of enzymes have evolved to stabilize the transition state(s) of their corresponding reactions \citep{casadevall2023alphafold2}. As a result, there is a geometric complementarity between the 3D shape of the protein active site and the molecular structure of the transition state. This complementarity determines to a large extent the catalytic activity of enzymes \citep{marti2004theoretical, liu2021large}. While we do not have access to ground truth or predicted transition states, we use the atom-mapping available in the  dataset to learn a superposition of the reactant and product molecular graphs and obtain the reaction embedding. 

Specifically, reactants and products are constructed as 2D graphs, where each molecular graph $\mathcal{G}=(\mathcal{V},\mathcal{E})$ has atom (node) features $v_i \in \mathcal{V} $ and bond (edge) features $e_{ij} \in \mathcal{E}$. A directed message-passing neural network (DMPNN) \citep{yang2019analyzing}, $f_{\text{mol}}$,  is used to separately encode the graph of the reactants $G_x$ and that of the products $G_y$. This results in learned atom and bond features $a_{i}, b_{ij} \in \mathbb{R}^d$. To simulate the transition state, we construct a pseudo-transition state graph, $G_{TS}=(\mathcal{V}_{TS},\mathcal{E}_{TS})$, by adding the bond features for edges connecting the same pairs of nodes in the reactants and the products. Edges between atom pairs that are not connected have bond features set to zero. We use the original atom features $v_i$ as the nodes of $\mathcal{G}_{TS}$ to preserve the atom identities. 
\begin{align}
    a_i, b_{ij} &= f_{\text{mol}}(\mathcal{G}_x, \mathcal{G}_y) \\ 
    v^{(TS)}_i &\coloneqq v^{(x)}_i \quad (=v^{(y)}_i) \\
    e^{(TS)}_{ij} &\coloneqq b_{ij}^{(x)} + b_{ij}^{(y)} 
\end{align}
We jointly train a second DMPNN, $f_{\text{TS}}$ to encode $G_{TS}$ and obtain the reaction embedding $r$ by aggregating the learned node features.
\begin{align}
    a'_i, b'_{ij} &= f_{\text{TS}}(\mathcal{G}_{TS}) \\ 
    \label{eq6} r &= \sum_i a'_i
\end{align}

\subsection{Protein Representation}
\label{sec:prot-enc}



Enzyme representation plays a pivotal role in modeling their function and interaction with substrates. To this end, we leverage advancements in both protein language models and graph neural networks. 

Each protein is represented as a 3D graph $\mathcal{G}_p=(\mathcal{V},\mathcal{E})$, with residue (node) features $h_i \in \mathcal{V} $ and bond (edge) features $e_{ij} \in \mathcal{E}$. Additionally each node $i$ has coordinates $c_i \in \mathbb{R}^3$. The node features of $\mathcal{G}_p$ are initialized using embeddings from the ESM-2 model with 650 million parameters (\verb|esm2_t33_650M_UR50D|) \citep{lin2022language}, which has demonstrated success in capturing many relevant protein features for a range of downstream tasks. The ESM model produces a feature vector for each residue denoted as $h \in \mathbb{R}^{1280}$.

To encode the protein graphs, we utilize an $E(n)$-Equivariant Graph Neural Network (EGNN) with coordinate updates \citep{satorras2021n}. This network is particularly suited for our purpose as it preserves translation, rotation and reflection equivariant graph features but is computationally inexpensive. Alternative methods preserve additional symmetries that are relevant to proteins such as SE(3) equivariance but are much more computationally expensive. We follow the implementation outlined in \citet{satorras2021n} except that the relative distances between nodes are encoded using a sinusoidal function (\cref{app:implementation}), as is common in protein structure modeling \citep{aykent2022gbpnet, atz2022delta, vaswani2017attention}.

\section{Experimental Setup}

\begin{figure*}[ht]
\vskip 0.1in
\begin{center}
\centerline{\includegraphics[width=\textwidth]{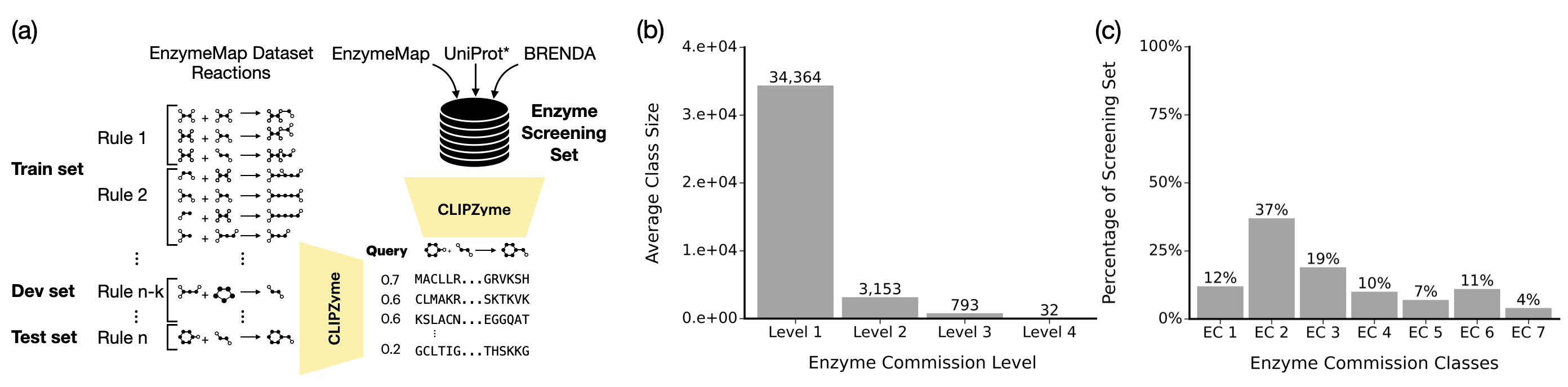}}
\caption{Overview of dataset construction and statistics. \textbf{(a)} Reaction-enzyme pairs are obtained from the EnzymeMap dataset \citep{heid2023enzymemap} and split based on their reaction rules. At test time, a reaction is queried and enzymes are ranked from a screening set composed of sequences from EnzymeMap, UniProt*, and BRENDA. 
\textbf{(b)} Averge number of sequences in each EC class when considering different levels of the EC hierarchy. \textbf{(c)} Distribution of sequences in the screening set according to their first EC level. \newline *The UniProt dataset is obtained from \citet{yu2023enzyme}.}
\label{fig:data}
\end{center}
\vskip -0.2in
\end{figure*}

\subsection{EnzymeMap Dataset}

Our method is developed on the EnzymeMap dataset \citep{heid2023enzymemap}, which includes biochemical reactions linked with associated UniProt IDs and their respective EC numbers. Each reaction is atom-mapped, allowing every atom in the product to be traced back to a corresponding atom in the reactants. To acquire the corresponding protein sequences, we select reactions linked to UniProt or SwissProt IDs and retrieve their sequences from their respective databases \citep{uniprot2023uniprot}. Additionally, we retrieve the predicted enzyme structures from the AlphaFold Protein Structure Database \citep{jumper2021highly, varadi2022alphafold}. We filter samples to include protein sequences up to 650 amino acids in length only. EnzymeMap provides a reaction rule for each reaction, which captures the bio-transformation in a reaction and can be applied to recreate the products of a reaction from its substrates \citep{ni2021curating}. With the goal of extending our model to unfamiliar chemical reactions, we divide our dataset into training, development, and testing groups based on these reaction rules (\cref{fig:data}). This results in a total of 46,356 enzyme-driven reactions, encompassing 16,776 distinct chemical reactions, 12,749 enzymes, across 2,841 EC numbers and 394 reaction rules.

EnzymeMap includes additional reactions that are associated with an EC number but lack an annotated protein sequence. We identify 7,967 of these unannotated reactions involving 1,101 EC numbers, distinct from our training data in terms of reaction rules. This subset serves as an additional validation set, allowing us to evaluate how our method ranks enzymes in relation to the EC number for each reaction. More information on how the data was processed can be found in \cref{app:data-prep}.

\subsection{Terpene Synthase Dataset}

Terpenoids are a large and diverse family of biomolecules with wide applications to medicine and consumer goods. The reactions generating these natural compounds involve particularly complex chemical transformations that are typically catalyzed by a class of enzymes called terpene synthases \citep{samusevich2023terpene}. This enzyme class is noteworthy for utilizing a relatively small number of substrates ($\sim$11) but is capable of generating thousands of distinct products. This presents a significant challenge with substantial implications. To further evaluate our method's performance on reactions known to involve challenging chemistry, we use a dataset of terpenoid reactions made available by recent work in detecting novel terpene synthases \citep{samusevich2023terpene}. We exclude reactions that are themselves or their enzyme included in our training set, obtaining 110 unique reactions and 99 enzymes.

\subsection{Enzyme Screening Set}

To construct our screening set of enzymes, we include sequences annotated in the EnzymeMap dataset \citep{heid2023enzymemap}, Brenda release \verb|2022_2|  \citep{chang2021brenda}, and those used in developing CLEAN (UniProt release \verb|2022_01|) \citep{yu2023enzyme}. We filter our set to those of sequence length $<650$ with available AlphaFold predicted structures \citep{jumper2021highly, varadi2022alphafold} and obtain a final list of 260,197 sequences.

\subsection{Baselines}

\subsubsection{Ranking Enzymes via EC Prediction}
\label{ec-compare}

The ultimate goal of enzyme screening is to identify candidate proteins from large protein databases, including the hundreds of millions of unannotated sequences. Since no standard computational procedure for enzyme screening has emerged, a reasonable approach is to first assign an EC number to the query reaction and then select all enzymes that share that EC class. To identify the EC classes of the enzymes in the screening set, one can use an EC predictor. On the other hand, assigning the full EC number of a reaction is not always straightforward or possible. For this reason, we consider baselines where between 1 to 4 levels of a query reaction's EC number are assignable (e.g., 1 level: 1.x.x.x to 4 levels: 1.2.3.4). We evaluate EC prediction and CLIPZyme on ranking the enzymes screening set for each reaction in the EnzymeMap test set.

\begin{figure}[h]
\vskip 0.1in
\begin{center}
\scalebox{0.85}{
\centerline{\includegraphics[width=\columnwidth]{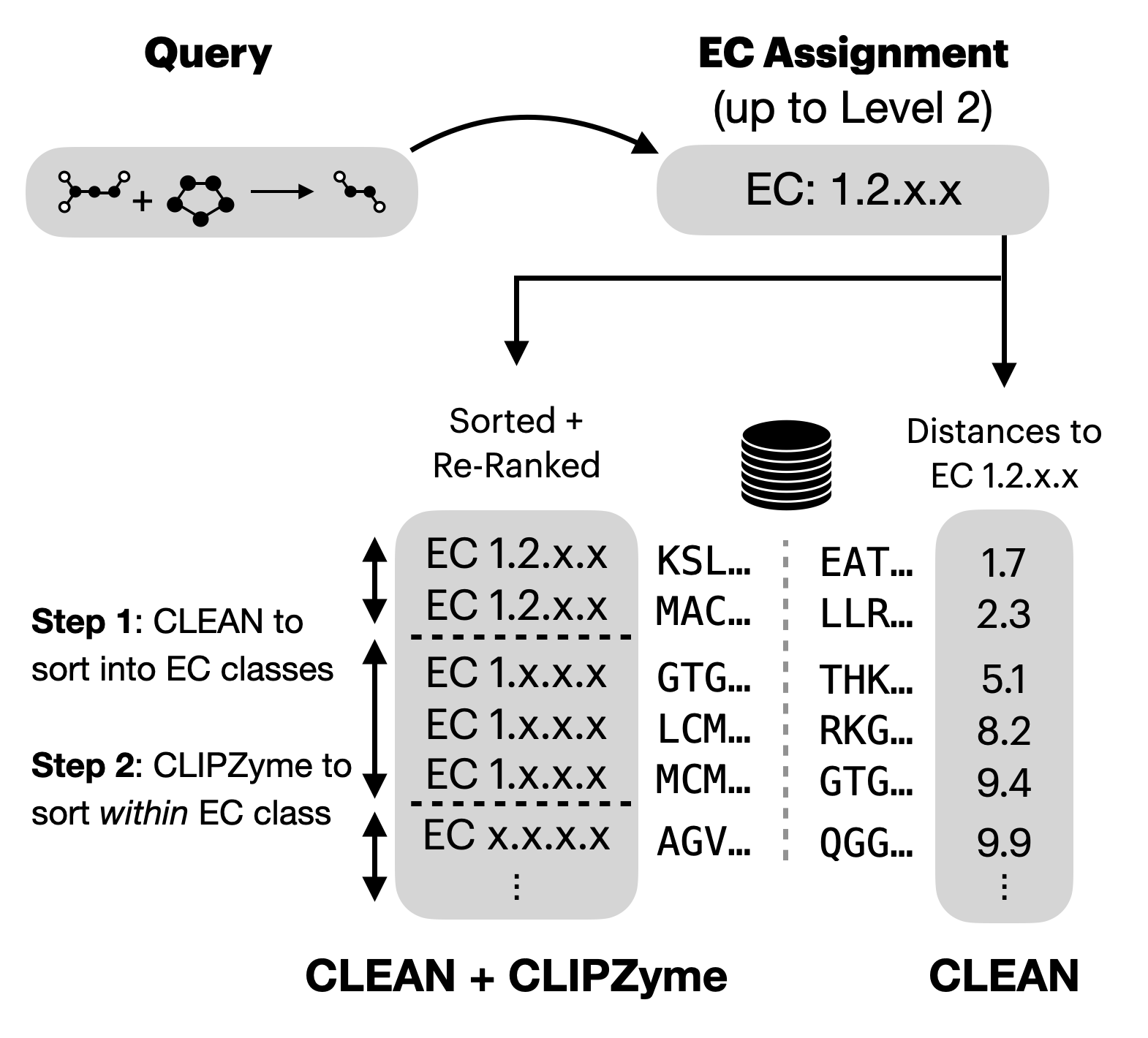}}}
\caption{Approaches for adapting EC prediction to virtual enzyme screening. We first assign a reaction an EC up to some level of specificity (here, level 2). To obtain rankings based on CLEAN, we use each sequence's distance to the EC class. To combine CLEAN and CLIPZyme, enzymes are first sorted according to their predicted EC class. Then they are ranked within each class using CLIPZyme.}
\label{fig:pipeline}
\end{center}
\vskip 0.1in
\end{figure}

We use CLEAN, a state-of-the-art EC predictor, to obtain a ranked list of enzymes for each EC \citep{yu2023enzyme}. CLEAN computes a single representation for every EC in its dataset as the mean embedding of sequences in that class and uses these as test-time anchors. The predicted EC class of a new sequence is then determined by the Euclidean distance to each EC anchor. Accordingly, given a reaction's assigned EC number, we rank our screening set enzymes by their distances to the reaction's EC anchor (\cref{fig:pipeline}). If a reaction's EC class does not exist in the CLEAN dataset, we broaden the search to one level higher. As an example, for a reaction with EC 1.2.3.4, if this EC is not in the CLEAN dataset, we rank enzymes according to their distances to the mean representation of EC 1.2.3 (and so on). For consistency with previous work, we use the CLEAN model trained on a split where none of the test enzymes share more than $50$\% sequence identity with those in the training set \citep{yu2023enzyme}.

We hypothesize that combining CLEAN to obtain EC predictions and CLIPZyme to rank them presents an opportunity for improved performance. Specifically, we predict the EC numbers for all of the enzyme sequences in our screening set using CLEAN. Given the reaction's assigned EC number, we first filter our screening set to those enzymes with the same exact predicted EC and rank this list using CLIPZyme (\cref{fig:pipeline}). We then filter all remaining enzymes to those that belong to one EC level above and again rank that list using CLIPZyme. As an example, given an input reaction with assigned EC of 1.2.3.4, we identify all enzymes predicted for that EC and rank them with CLIPZyme. We then rank all remaining enzymes with predicted EC 1.2.3. This process is repeated until all enzymes are ranked.

\subsubsection{Reaction Representation}
We explore three alternative methods for encoding the reaction and compare against these in our results. The first uses the condensed graph reaction (CGR) representation \citep{hoonakker2011condensed} by overlaying the reactants and products and concatenating the edge features. A DMPNN encodes the CGR to obtain a hidden representation of the reaction. The second approach is to use the full reaction SMILES \citep{weininger1988smiles} as an input to a language model and obtain a final representation of the reaction. We follow the tokenization scheme for SMILES introduced by the Molecular Transformer \citep{schwaller2019molecular} and train a transformer model  as our encoder \citep{vaswani2017attention}. We also consider the Weisfeiler-Lehman Difference Network (WLDN) architecture and implement it as described in \citet{jin2017predicting}. We train all models until convergence, using the same data splits and hyper-parameters (\cref{app:training}).

\subsubsection{Protein Representation}
We focus on achieving a balance between efficiency and the ability to process extensive enzyme datasets. To this end, we explore both sequence-based and structure-based approaches, acknowledging the critical influence of structure on enzymatic activity despite its inherent computational demands. We train ESM-2 \citep{lin2022language} as a sequence-based baseline for protein encoding. We also encode the structure with an EGNN \citep{satorras2021n} and compare initializing node embeddings from either the MSA-transformer \citep{rao2021msa} or ESM-2, to identify the best method in terms of both performance and speed.


\begin{table*}[!ht]
\caption{Enzyme virtual screening performance compared to using EC prediction alone and together with CLIPZyme. For a given reaction EC level, enzymes are ranked according to their Euclidean distance to EC class anchors when using CLEAN \citep{yu2023enzyme}. Alternatively, CLEAN is first used to place enzymes into successively broader EC levels matching that of the reaction, and CLIPZyme is used to reorder the enzymes within each level. BEDROC: Boltzmann-enhanced discrimination of receiver operating characteristic; EF: enrichment factor.}
\label{tab:clean}
\vskip 0.15in
\begin{center}
\begin{small}
\begin{sc}
\begin{tabular}{llcccr}
\toprule
EC Level Assumed Available & Method & BEDROC$_{85}$  (\%)& BEDROC$_{20}$  (\%)& EF$_{0.05}$ & EF$_{0.1}$   \\
\midrule
- & CLIPZyme & 44.69 & 62.98 & 14.09 & 8.06  \\ 
\hline 
\multirow{2}{*}{Level 1 (x.-.-.-)} & CLEAN  & 0.96& 6.53& 1.22& 1.72\\  \vspace{0.5em}
 & CLIPZyme + CLEAN  & 57.03& 78.50& 17.84& 9.56\\ 
\multirow{2}{*}{Level 2 (x.x.-.-)} & CLEAN  & 4.86& 14.10& 3.23& 2.49\\ \vspace{0.5em}
& CLIPZyme + CLEAN  & 75.57& 90.20& 19.40& 9.84\\
\multirow{2}{*}{Level 3 (x.x.x.-)} & CLEAN  & 25.86& 36.75& 8.03& 4.81\\ \vspace{0.5em}
& CLIPZyme + CLEAN  &82.69& 93.23&  19.43& 9.84\\
\multirow{2}{*}{Level 4 (x.x.x.x)} & CLEAN   &89.74&93.42& 18.97& 9.60\\
& CLIPZyme + CLEAN  & 89.57& 95.24& 19.43& 9.84\\
\bottomrule
\end{tabular}
\end{sc}
\end{small}
\end{center}
\vskip -0.1in
\end{table*}

\subsection{Evaluation Setup}

We aim to simulate the scenario where an enzyme is desired to catalyze a novel reaction, and it exists in nature but is not annotated. We compare different approaches to encoding the reactions and their enzymes, and compare our method to an alternative approach using  EC prediction. 

As our main aim is to generalize to novel chemical transformations, our test set consists of reactions with reaction rules that are unseen during training, queried against all 260,197 sequences. However, this means our screening set does include proteins used in training the model. Therefore, we also evaluate model performance when excluding proteins used in training. Additionally, we use MMSeqs2 \citep{steinegger2017mmseqs2} and Foldseek \citep{van2023fast} to exclude proteins based on their similarity to the training set proteins in terms of sequence identity and protein fold, respectively. If the exclusion of a protein results in a test reaction having no actives in the screening set, we exclude the entire reaction. 

Throughout our evaluations, we take the BEDROC score as our primary metric \citep{truchon2007evaluating}. We focus on the case $\alpha=85$, where the top 3.5\% of predictions contribute to 95\% of the score, and as suggested in \citet{truchon2007evaluating}, we also calculate the BEDROC score for $\alpha=20$. We also report the enrichment factor (EF) when taking the top 0.5\% and 1\% of predictions. This estimates the fraction of catalyzing enzymes found in our top predictions relative to random selection.


\section{Results}

We present here an overview of our key results. In \cref{tab:clean}, we compare CLIPZyme's performance to that of EC prediction with CLEAN and show the benefit of combining methods. CLIPZyme shows improved performance in all comparisons. \cref{tab:ablation} shows the impact of different protein and reaction representations and highlights the superior performance of our novel reaction encoding. In \cref{tab:rxn-specific}, we show that CLIPZyme's performance extends to a challenging dataset of terpene synthase reactions and unannotated reactions. Lastly, we show in \cref{tab:exclusion} that CLIPZyme's performance drops when screening enzymes that significantly differ from those it was trained on, but still maintains useful predictive value. Additional analysis is provided in \cref{app:analysis}.

\subsection{Enzyme Screening Evaluation on EnzymeMap}
\label{subsec:ec-pred}

CLIPZyme effectively ranks the screening set against reactions in the EnzymeMap test set with an average BEDROC$_{85}$ of 44.69\% and an enrichment factor of 14.09 when choosing the top 5\% (\cref{tab:clean}). We compare its performance to ranking using EC prediction with CLEAN. Since it is not always possible to assign all 4 levels of an EC to a chemical reaction, we examine scenarios where different EC levels are assumed to be known for query reactions in the test set. 

For example, with only the first EC level known, using EC prediction alone obtains a BEDROC$_{85}$ score of 0.96\% (\cref{tab:clean}). This improves to 25.86\% when we are able to specify a reaction up to the third EC level. With four EC levels known, the CLEAN method becomes more effective than CLIPZyme alone. However, being able to assign all four EC levels for a reaction may not be always feasible in real-world applications.

Combining the CLEAN method with CLIPZyme achieves improved performance regardless of how many EC levels we assume to be known for reactions. Here, CLEAN is first used to predict the EC classes of enzymes in the screening set. Enzymes within the predicted EC class are re-ranked using CLIPZyme (\cref{fig:pipeline}). Even basic knowledge of the first EC level of a chemical reaction enhances CLIPZyme's performance from a BEDROC$_{85}$ of 44.69\% to 57.03\%. With the first two levels assumed to be known, performance also improves to 75.57\%.


We note that EC classification may be insufficient for categorizing chemical reactions that do not fit in existing EC classes. As a result, any EC prediction method is not applicable in that setting, while CLIPZyme is as it operates directly on the reaction. 

\begin{table*}[!ht]
\caption{Performance of various protein and reaction encoding schemes on virtual screening for reactions in the EnzymeMap test set. The symbol {\SnowflakeChevron} denotes models where the weights are kept unchanged during training.}
\label{tab:ablation}
\vskip 0.15in
\centering
\begin{small}
\begin{sc}
\scalebox{0.95}{
\begin{tabular}{llcccr}
\toprule
Protein Encoder & Reaction  Encoder & BEDROC$_{85}$(\%)	& BEDROC$_{20}$(\%) & EF$_{0.05}$ & EF$_{0.1}$ \\
\midrule
ESM\textsuperscript{\SnowflakeChevron} & Ours (\cref{sec:rxn-enc}) & 17.84  & 29.39  & 6.61 & 4.17 \\
ESM & Ours (\cref{sec:rxn-enc}) & 36.91  & 53.04  & 11.93 & 6.84 \\
MSA-Transformer\textsuperscript{\SnowflakeChevron} + EGNN 
    & Ours (\cref{sec:rxn-enc}) & 28.76  & 46.53 & 10.34 & 6.67    \\
ESM\textsuperscript{\SnowflakeChevron} + EGNN 
    & CGR \citep{hoonakker2011condensed} & 38.91 & 57.58  & 13.16  & 7.73  \\
ESM\textsuperscript{\SnowflakeChevron} + EGNN 
    & Reaction SMILES & 29.94 & 46.01  & 10.34 & 6.32 \\
 ESM\textsuperscript{\SnowflakeChevron} + EGNN & WLDN \citep{jin2017predicting} & 29.84& 46.70& 10.71&6.41\\
ESM\textsuperscript{\SnowflakeChevron}  + EGNN 
    & Ours (\cref{sec:rxn-enc}) & 44.69 & 62.98 & 14.09 & 8.06  \\
\bottomrule
\end{tabular}}
\end{sc}
\end{small}
\vskip -0.1in
\end{table*}

\subsection{Impact of Reaction and Protein Representation}
\label{subsec:ablation}

We explore a number of different encoding methods for both reaction and protein representations and find that the model is highly sensitive to changes in both (\cref{tab:ablation}). Using the molecular structures of the reaction obtains better performance than language-based methods operating over the reaction SMILES, with the former achieving a BEDROC$_{85}$ of 44.69\% compared to 29.94\%. This suggests that structural representations may capture chemical transformations that correspond to enzyme activity more explicitly than language based ones. The patterns observed in structures may be more difficult for language models to capture without additional features or data. Employing a more expressive model also improves performance when compared to using WLDN as the reaction encoder. While all reaction representation methods include the full reaction, they differ in how the bond changes are encoded. Methods that explicitly delineate chemical transformations between substrates and products appear to obtain generally better performance.

We find a similar sensitivity to enzyme encoding. We compare using ESM embeddings alone and using ESM embeddings together as node features for EGNN. We find that using an EGNN to capture the structural components of the enzyme improves performance compared to training a sequence-based model alone (44.69\% compared to 36.91\%), which indicates that enzyme structure is important for achieving good performance on this task. We also explore initializing the EGNN node features with embeddings from the pre-trained MSA-Transformer \citep{rao2021msa}. These embeddings do not appear to improve performance, although they capture evolutionary information of the sequence. This, however, may be due to differences in quality of representations learned by ESM and MSA-Transformer in which ESM-2 was trained on much larger set of sequences.

\begin{table}[ht]
\vskip -0.1in
\caption{Performance of CLIPZyme on additional biochemical reactions. The terpene synthase dataset is obtained from \citet{samusevich2023terpene} and includes reactions considered to involve more complex biotransformations. The unannotated subset of EnzymeMap consists of reactions in the dataset that are not assigned a UniProt or SwissProt identifier. In this case, virtual screening is evaluated as the ability to highly rank proteins with the correct EC class. }
\label{tab:rxn-specific}
\begin{center}
\begin{small}
\begin{sc}
\scalebox{0.8}{
\begin{tabular}{m{6em}cccc}
\toprule
Dataset &  BEDROC$_{85}$  (\%)& BEDROC$_{20}$  (\%)& EF$_{0.05}$ & EF$_{0.1}$ \\
\midrule
Terpene Synthases & 72.46& 85.89&18.29 & 9.42\\
\hline
Unannotated EnzymeMap & 42.94 & 61.39 &13.92 & 7.73 \\
\bottomrule
\end{tabular}}
\end{sc}
\end{small}
\end{center}
\end{table}

\subsection{Evaluation on Reaction-Specific Datasets}
\label{subsec:rxn-specific}
We extend our evaluation to two additional datasets to further assess CLIPZyme's utility in practical applications in \cref{tab:rxn-specific}. The first dataset encompasses reactions catalyzed by terpene synthases. We evaluated CLIPZyme using the same screening set and observed robust performance, evidenced by a BEDROC$_{85}$ score of 72.45\%. Due to the small and uniform substrate pool, the model might be preferentially ranking terpene synthases as a whole, rather than effectively distinguishing between specific reactions.

Additionally, we present an evaluation using unannotated reactions from EnzymeMap. For the sake of evaluation, we assume the true enzymes in the screening set for a given reaction are those with EC classes matching that of the reaction. Under this setup, CLIPZyme achieves a BEDROC$_{85}$ of 42.94\%, which aligns closely with the results from the annotated subset of EnzymeMap. Because the metrics are calculated relative to the EC classes of each protein, this result suggests that the CLIPZyme rankings correspond with the proteins' EC numbers.

\begin{table*}[t]
\caption{Performance when excluding sequences from the screening set with various levels of similarity to training set enzymes.}
\label{tab:exclusion}
\vskip 0.15in
\begin{center}
\begin{small}
\begin{sc}
\begin{tabular}{lcccc}
\toprule
Exclusion criteria & BEDROC$_{85}$  (\%)& BEDROC$_{20}$  (\%)& EF$_{0.05}$ & EF$_{0.1}$ \\
\midrule
Exact Match & 39.13 & 58.86 & 13.40 & 7.81   \\ 
MMSeqs 30\% Similarity & 35.32& 54.86&12.43& 7.30\\ 
Foldseek 30\% Similarity & 21.44& 35.39& 7.93& 4.93\\ 
\bottomrule
\end{tabular}
\end{sc}
\end{small}
\end{center}
\vskip -0.1in
\end{table*}

\subsection{Generalization to Novel Proteins}
\label{subsec:generalization}
Our primary focus has been on evaluating the generalization of CLIPZyme on  reactions unseen during training. However, given the ultimate goal of screening a wide array of both annotated and unannotated enzymes, it's crucial to understand the model's efficacy in ranking proteins dissimilar to those in the training set.

To do so, we exclude proteins that are similar to our training set according to three similarity metrics. We first exclude training set enzymes. Second, we apply MMSeqs2 \citep{steinegger2017mmseqs2} to remove enzymes with 30\% or greater sequence similarity. Lastly, we exclude enzymes with 30\% fold similarity as determined by Foldseek \citep{van2023fast}. By measuring performance on these three screening subsets, we demonstrate CLIPZyme's generalizability across both reactions and enzymes.

Each exclusion criteria led to a  reduction in performance. For example, CLIPZyme's performance decreases by approximately 5 percentage points on both BEDROC metrics when excluding 
training set enzymes \cref{tab:exclusion}. The most marked impact was observed with Foldseek-based filtering, showing a 23.25 point decrease in BEDROC$_{85}$ scores. This aligns with our previous findings that protein structural features play a critical role in effective screening. Despite this, the model still demonstrated a notable ability to rank enzymes effectively as the top-ranked candidates consistently showed enrichment for active enzymes.

\section{Conclusion}
We present here the task of virtual enzyme screening and a contrastive method, CLIPZyme, to address it. We show that our method can preferentially rank catalytically active enzymes against reactions across multiple datasets. Without a standard baseline, we examine enzyme screening through EC prediction and highlight CLIPZyme's competitive ability. We futhermore show that combining EC prediction with CLIPZyme  achieves significantly improved performance. Lastly, we evaluate CLIPZyme's capacity to generalize by evaluating it on additional challenging reaction datasets and on unseen protein clusters. In practical scenarios, where millions or even hundreds of millions of enzymes need screening, we foresee the necessity of methods like CLIPZyme with even higher sensitivity for effective enzyme design at scale. 

Among its limitations, the current approach does not model the physical interactions between reactants and enzymes, and it is unable to capture the mechanisms that give rise to the observed reaction. Moreover, the available data covers a relatively small chemical space and includes a restricted set of reactions and enzyme sequences (e.g., EC class 7 is completely unrepresented). We also note that our approach of random negative sampling may give rise to false negatives due to the promiscuity of many enzymes and the method may benefit from alternative sampling techniques.  Directions for future work include modeling the 3D interactions characterizing biochemical reactions (e.g., through docking) and leveraging transition state sampling through quantum chemical simulations.

\section*{Impact Statement}
The ability to identify, repurpose, or create enzymes for novel reactions  remains a grand challenge that has profound societal impact through applications ranging from therapeutic manufacturing to the biodegradation of plastics. By enriching the pool of candidate enzymes through virtual screening, experimental approaches to optimize enzymes gain a greater chance of achieving efficient catalysis. As with all computational methods that facilitate the generation of small molecules, successful virtual enzyme screening may also be exploited for the production of hazardous chemicals.

\section*{Software and Data}
We download EnzymeMap from \url{https://github.com/hesther/enzymemap}, and the Terpene Synthase data from \url{https://zenodo.org/records/10359046}.

\section*{Acknowledgements}
This work is supported by the Jameel Clinic for AI and Health at MIT, the Eric and Wendy Schmidt Center at the Broad Institute of MIT and Harvard, and Novo Nordisk A/S. We are also grateful for Jeremy Wohlwend, Anush Chiappino-Pepe, Tomáš Pluskal, Raman Samusevich, Esther Heid, Hannes Stark, and Bowen Jing for their insightful discussions.

\bibliography{example_paper}
\bibliographystyle{icml2024}

\newpage
\appendix
\onecolumn

\section{Additional Analysis}
\label{app:analysis}
\subsection{Enzyme Screening Within EC Classes}
\label{subsec:within-ec}

\begin{table}[h]
\vskip -0.1in
\centering
\caption{Performance of CLIPZyme when limiting the screening set to enzymes belonging to the query reaction's top EC level.}
\begin{center}
\begin{small}
\begin{sc}
\begin{tabular}{lcccc}
\toprule
 BEDROC$_{85}$ (\%)& BEDROC$_{20}$  (\%)& EF$_{0.05}$ & EF$_{0.1}$ \\
\midrule
 36.25&  51.61&  11.30&  6.83\\ 
 \bottomrule
\end{tabular}
\end{sc}
\end{small}
\end{center}
\label{tab:within-ec}
\vskip -0.1in
\end{table}

We also explore CLIPZyme's ability to discriminate between enzymes within the same EC class, where enzymes are more likely to share function and physical-chemical features. To do so, for each query reaction in the test set, we adjust the screening set to include only those enzymes belonging to its EC class. The number of enzymes quickly diminishes when considering EC subclasses  to the extent that the EC-based screening sets become too small for virtual screening (\cref{fig:data}b) -- for example, the BEDROC metric is only valid only when $(\alpha \times \text{proportion of actives}) \ll 1$. For this reason, we consider only the top EC level in this analysis. We observe that it is more difficult to rank the correct enzymes higher when only considering sequences in the same EC class but that the top predictions are still enriched for the active enzymes (\cref{tab:within-ec}).

\subsection{Adapting CLEAN for Ranking Enzymes}
\label{app:training}

We consider using both CLEAN EC predictions and computed distances to perform virtual screening similar to \cref{ec-compare}. Here we present an alternative reranking approach than that in the main body. We follow the exact same setup as reranking EC predictions using CLIPZyme but instead rerank using the distance to the EC anchors. For example, given a query reaction with EC 1.2.3.4, we first predict the EC numbers for all of the enzymes in the screening set using CLEAN. We then rank the enzymes with predicted EC of 1.2.3.4 by the distance from the anchor with EC 1.2.3.4 (computed as the mean embedding of all ECs in the CLEAN training set with EC 1.2.3.4). We then rank all remaining enzymes with predicted EC of 1.2.3.x by their distance to the anchor embeddings of EC 1.2.3.4 (this is the same anchor). This differs from the main body approach since CLEAN assigns EC numbers based on a varying threshold (i.e., max-separation) for each embedding. By first ordering by EC and then reranking within each EC we achieve different results than by ranking all at once by distance to the 1.2.3.4 anchor. 

\begin{table*}[h]
\caption{Enzyme virtual screening performance when using CLEAN to first place enzymes into successively broader EC levels matching that of the reaction, then re-ranking them according to their Euclidean distance to the reaction’s EC.}
\label{tab:clean2}
\vskip 0.15in
\begin{center}
\begin{small}
\begin{sc}
\begin{tabular}{lcccr}
\toprule
EC Level & BEDROC$_{85}$  (\%)& BEDROC$_{20}$  (\%)& EF$_{0.05}$ & EF$_{0.1}$   \\  
\midrule
Level 1& 5.43 & 26.94 & 5.55 & 6.33\\ 
Level 2& 35.56& 71.10& 18.95& 9.72\\ 
Level 3& 63.40 & 85.61 & 19.35  & 9.74\\
Level 4& 92.65& 96.16& 19.48& 9.80\\
\bottomrule
\end{tabular}
\end{sc}
\end{small}
\end{center}
\vskip -0.1in
\end{table*}

\section{Data Processing}
\label{app:data-prep}

\subsection{Enzymemap}
\label{app:emap}
We obtain version 2 of the EnzymeMap dataset \citep{heid2023enzymemap} and use only the reactions with assigned protein references from either SwissProt or UniProt. Our method requires that the same atoms appear on both sides of the reaction, so we exclude samples where this is not the case. We also filter reactions where the EC number is not fully specified, the sequence could not be retrieved from UniProt, or there wasn't a computable bond change. We restrict our data to proteins of sequence length no more than 650 (maintaining 90\% of the sequences) and those with a predicted structure in the AlphaFold database. We remove duplicate reactions, where the same reaction and sequence appear for multiple organisms. We split reactions into train/development/test splits with a ratio of 0.8/0.1/0.1 based on the reaction rule IDs assigned in the dataset. The statistics for the final dataset are shown in \cref{tab:data-stat}.

\begin{table}[h]
    \caption{Statistics of the EnzymeMap dataset used to develop CLIPZyme after pre-processing.}
    \label{tab:data-stat}
    \vskip 0.15in
    \begin{center}
    \begin{small}
    \begin{sc}
    \begin{tabular}{lccc}
    \toprule
         &  Training Split&  Development Split& Test Split\\
         \hline
         Number of Samples&  34,427&  7,287& 4,642\\
         Number of Reactions&  12,629&  2,669& 1,554\\
         Number of proteins&  9,794&  1,964& 1,407\\
         Number of ECs&  2,251&  465& 319\\
         \bottomrule
    \end{tabular}
    \end{sc}
    \end{small}
    \end{center}
\end{table}

\subsection{Protein Structures}
\label{app:cifs}
We obtain all protein structures as CIF files from the AlphaFold Protein Structure Database \citep{jumper2021highly, varadi2022alphafold}. We parse these files using the BioPython MMCIFParser. We then construct graphs for use in the PyTorch Geometric library \citep{fay2019pyg}. First we filter out the atoms from the CIF file to only include the $C_\alpha$ atoms of the protein. Each graph node as a result represents a residue and the associated coordinates from the CIF file. The edges are determined using the k-nearest neighbors (kNN) method, creating a connected graph that reflects the chemical interactions within the protein. We use a distance of 10 angstroms as a cutoff for the edges.

\subsection{MSA Embeddings}
\label{app:msa}
We explore using the hidden representations from the MSA Transformer \citep{rao2021msa} as node embeddings of the enzyme 3D structure. Rather than using HHblits \citep{remmert2012hhblits}, we opt for MMSeqs2 \citep{steinegger2017mmseqs2} because of its speed and efficient search. We follow the pipeline employed by ColabFold \citep{mirdita2022colabfold} but use only the UniRef30 (\verb|uniref30_2302|) database and do not use an expanded search \citep{suzek2015uniref, mirdita2017uniclust}. We sample 128 sequences for each MSA using a greedy search (maximum similarity) to obtain the input for the MSA-Transformer. We keep only the hidden representations of the query enzyme sequence and discard those from the MSA search. For an enzyme of length $n$, this yields sequence embeddings $h \in \mathbb{R}^{n \times 768}$.

\subsection{Computing Screening Set Enzyme Clusters }

To exclude from our enzyme screening set those proteins that are similar to sequences used in our training dataset, we compute protein clusters using MMSeqs2 \citep{steinegger2017mmseqs2} and Foldseek \citep{van2023fast}. For MMSeqs2, we use the default parameters with \verb|--min-seq-id|$=0.3$ and \verb|--similarity|$=0.8$. For Foldseek, we use the default parameters with \verb|--min-seq-id|$=0$ and \verb|--c|$=0.3$.

\section{Implementation Details}
\label{app:implementation}

All models are developed in PyTorch v2.0.1 \citep{Paszke_PyTorch_An_Imperative_2019} and trained using PyTorch Lightning v2.0.9 \citep{Falcon_PyTorch_Lightning_2019}.

\paragraph{$f_{\text{mol}}$ and $f_{\text{TS}}$} We implement our reaction encoder (\ref{sec:rxn-enc}) as two DMPNNs \citep{yang2019analyzing}. We use standard node and edge features (\cref{tab:mol-feats}) to initialize the reactant and product graphs, with input node dimensions of 9 and input edge  dimensions of 3. The first encoder, $f_{\text{mol}}$ has 5 layers and a hidden dimension of 1,280. The node features for the second encoder, $f_{\text{TS}}$ are unchanged, while edges are obtained from taking the sum of the hidden edge representations from $f_{\text{mol}}$. Hence the node dimensions are still 9, while the input edge features have dimensions 1,280. The model also consists of 5 layers and a hidden size of 1,280. We aggregate the graph as a sum over the node features.

\begin{table}[h]
    \caption{Chemical properties used as node and edge features in constructing molecular graphs.}
    \label{tab:mol-feats}
    \vskip 0.15in
    \begin{center}
    \begin{small}
    \begin{sc}
    \begin{tabular}{l p{8cm}}
    \toprule
         Entity &  Features \\
         \hline
          atom (node) features&  atomic number, chirality, degree, formal charge, number of hydrogens, number of radical electrons, hybridization, aromaticity,  belonging to a ring\\
         bond (edge) features&  bond type, stereochemistry, conjugation\\
         \bottomrule
    \end{tabular}
    \end{sc}
    \end{small}
    \end{center}
\end{table}

\paragraph{Condensed Graph Reaction}
We construct the condensed graph reaction as described in \citet{heid2021machine}. Specifically, the atom and edge features for the reactants and products are created as binary vectors for the properties detailed in \cref{tab:mol-feats}. For node features $x_i^{(r)}, x_i^{(p)}$ and edge features  $e_{ij}^{(r)}, e_{ij}^{(p)}$, we compute $x' = x_i^{(r)} - x_i^{(p)}$ and $e_{ij}' = e_{ij}^{(r)} -  e_{ij}^{(p)}$. We do not use the atomic number in calculating $x'$. Concatenating these with our reactants' features, our final CGR graph consists of 225 atom and 26 edge features, $x_i^{CGR} = [x_i^{(r)} \mathbin\Vert x_i']$ and $ e_{ij}^{CGR} = [e^{(r)}_{ij} \mathbin\Vert e'_{ij} ]$, respectively. 

\paragraph{Reaction SMILES} The reaction SMILES is first canonicalized then tokenized according to \citet{schwaller2019molecular} without atom-mapping. We create a vocabulary based on this tokenization scheme and use a tranformer architecture \citep{vaswani2017attention} as implemented by the Hugging Face library (we use the BertModel) \citep{wolf-etal-2020-transformers}. The transformer is initialized with 4 layers, a hidden and intermediate size of 1,280, and 16 attention heads. An absolute positional encoding is used over a maximum sequence length of 1,000. We prepend the reaction with a \verb|[CLS]| token and use its hidden representation as the reaction embedding.

\paragraph{WLDN} We implement WLDN as originally described in \citet{jin2017predicting} and initialize it with 5 layers and a hidden dimension of 1,280. The difference graph is calculated as the difference between atom-mapped node embeddings of the substrate and product graph. We apply a separate 1-layer WLN to obtain the final graph-level representation.

\paragraph{EGNN} Node features are initialized with residue-level embeddings from ESM-2 (the 650M parameter variant with 33 layers) \citep{lin2022language}. We use a hidden size of 1,280, 6 layers, and a message dimension of 24. Both features and coordinates are normalized and updated at each step. Neighborhood aggregation is done as an average, and protein-level features are taken as a sum over the final node embeddings. Repurposing the positional encodings used in \citet{vaswani2017attention}, pairwise distances are transformed with sinusoidal embeddings. For a given relative distance $d_{ij}$ between nodes $i$ and $j$, the encoding function $f : \mathbb{N} \rightarrow \mathbb{R}^d$ transforms this distance into a $d$-dimensional sinusoidal embedding. The encoding is defined as follows:
\begin{equation}
    f(d_{ij})^{(k)} = 
    \begin{cases} 
        \sin\left(\frac{1}{\theta^{k/2}} \cdot d_{ij}\right), k < \frac{d}{2}, \\
        \cos\left(\frac{1}{\theta^{\frac{k-\frac{d}{2}}{2}}} \cdot d_{ij}\right), k \geq \frac{d}{2}. 
    \end{cases} 
\end{equation}
where $k$ is the index of the dimension of the distance vector, $\theta$ is a hyperparameter that controls the frequency of the sinusoids, which in our case is set to 10,000. The resulting embedding for a particular relative distance $d_{ij}$ is constructed by concatenating the sine-encoded and cosine-encoded vectors, thus interleaving sinusoidal functions along the dimensionality of the embedding space.

\paragraph{CLEAN} We train CLEAN with the supervised contrastive ("Supcon-Hard") loss following the training protocol and parameters loss described in the project's repository (\url{https://github.com/tttianhao/CLEAN}). Specifically, we use the supervised contrastive loss and the data split in which none of the test enzymes share $> 50\% $ sequence identity with those in the training set.  At inference, we use the same approach described in \citet{yu2023enzyme} to compute the EC anchors. We obtain the predicted distance between each enzyme in our screening set and each EC anchor. We extend this to parent classes of the ECs. For instance, the representation for EC 1.2.3.x is the mean embedding of all CLEAN proteins in that class. We also predict the EC numbers for all of the enzyme sequences in our screening set using the ``max-separation" algorithm.

\section{Training Details}
\label{app:training}
All models are trained with a batch size of 64 with bfloat16 precision and trained until convergence (approximately 30 epochs). We use a learning rate of $1e^{-4}$ with a cosine learning rate schedule and 100 steps of linear warm-up. Warm-up starts with a learning rate of $1e^{-6}$, and the minimum learning rate after warm-up is set to $1e^{-5}$. We use the AdamW optimizer \citep{loshchilov2017decoupled} with a weight decay of 0.05 and ($\beta_1, \beta_2) = (0.9, 0.999)$. When training the ESM model, we initialize with the pretrained weights of \verb|esm2_t33_650M_UR50D| and use a mean of the residue embeddings for the sequence representation. We train all models on 8 NVIDIA A6000 GPUs.

\end{document}